\documentclass{PoS}

\usepackage[utf8]{inputenc}
\usepackage{natbib}
\usepackage{amssymb}
\usepackage{fontenc}
\usepackage{times}
\usepackage{mathptmx}
\usepackage{graphicx}
\usepackage{pdfpages}
\usepackage{verbatim}
\usepackage{float}
\usepackage{amsmath}
\usepackage{color}
\usepackage{tabularx}
\usepackage{longtable}
\usepackage{wrapfig}
\usepackage{subfig}

\title{Mass Assembly of High-z Galaxies with MASSIV}

\ShortTitle{Mass Assembly of High-z Galaxies with MASSIV}

\author{\speaker{V. Perret},  B. Epinat, P. Amram, O. Le Fèvre, C. Lopez-Sanjuan, L. Tasca, L. Tresse\\
        Aix Marseille Université, CNRS, LAM (Laboratoire d'Astrophysique de Marseille) UMR 7326, 13388, Marseille, France\\
        E-mail: \email{valentin.perret@oamp.fr}}
\author{T. Contini, C. Divoy, J. Queyrel, J. Moultaka\\
	IRAP, Université de Toulouse, UPS-OMP, CNRS, Toulouse, France
}
\author{D. Vergani\\
	INAF IASFBO, Via P. Gobetti 101, I-40129 Bologna, Italy
}
\author{F. Bournaud\\
	CEA, IRFU, SAp, F-91191 Gif-sur-Yvette, France
}
\author{B. Garilli, L. Paioro\\
	INAF-IASFMI, Via E. Bassini 15, I-20133 Milano, Italy
}
\author{M. Kissler-Patig\\
	ESO, Karl-Schwarzschild-Str. 2, D-85748 Garching b. München, Germany
}

\abstract{MASSIV (Mass Assembly Survey with SINFONI in VVDS) is a sample of 84 distant star-forming galaxies observed with the SINFONI Integral Field Unit (IFU) on the VLT. These galaxies are selected inside a redshift range of 0.8 < z < 1.9, i.e. where they are between 3 and 5 billion years old. The sample aims to probe the dynamical and chemical abundances properties of representative galaxies of this cosmological era. On the one hand, close environment study shows that about a third of the sample is involved in major mergers. On the other hand, kinematical analysis revealed that 42\% of the sample is rotating disks, in accordance with higher redshift samples. The remaining 58\% show complex kinematics, suggesting a dynamical support based on dispersion, and about half of these galaxies is involved in major mergers.  Spheroids, unrelaxed merger remnants, or extremely turbulent disks might be an explanation for such a behavior. Furthermore, the spatially resolved metallicity analysis reveals positive gradients, adding a piece to the puzzle of galaxies evolution scenarios.}

\FullConference{VIII International Workshop on the Dark Side of the Universe,\\
		June 10-15, 2012\\
		Rio de Janeiro, Brazil}

\begin{document}
\bibliographystyle{aa}
\section{Mass Assembly of High Redshift Galaxies}

In the current framework of $\Lambda$CDM cosmology, the mean energy density of the Universe is shared between dark energy, dark matter and baryonic matter. In this context, the formation of structures in the Universe is seeded by small perturbations in matter density expanded to cosmological scales by inflation. Given such a cosmological model, the Universe is about 13.7 billion years old, and galaxies are expected to form around z=20-50 \citep{2007MNRAS.378..449G}, when the first sufficiently deep dark matter potential wells are made and allowed gas to cool and condense to produce primeval stars and galaxies.

This standard model reproduces well the linear initial conditions, the intergalactic medium structure during galaxy formation, and large scale structure as observed today. However, the hierarchical dark matter halo formation paradigm remains strongly debated. The physical processes responsible for mass assembly at early epochs is still unclear. While merging events are expected to play an important role in the building of the  Hubble sequence, smooth cold gas accretion might also have strongly contributed to the growth of galaxies. 
Secular processes such as stellar feedback may also have a major role in the build-up of local galaxies. High-z galaxies tend to have a higher gas fraction, thus a powerful stellar feedback would be able to drive the evolution within the gravitational potential.

\begin{wrapfigure}{r}{0.7\textwidth}
	\begin{center}
		\includegraphics[trim=0cm 0.5cm 0cm 1cm,width=0.70\textwidth]{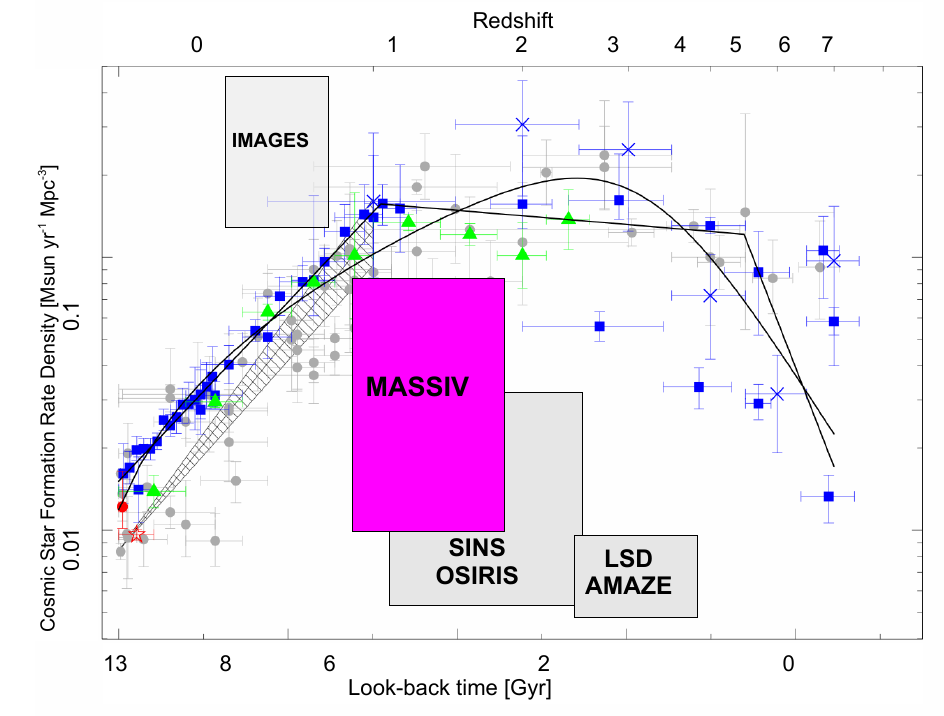}
	\end{center}
	\caption{Evolution of the cosmic star formation rate density as a function of look-back time and redshift. Major IFU surveys redshift ranges are compared. The relative height of each box is proportional to the sample size.  Adapted from \cite{2012A&amp;A...539A..91C}.}
	\label{cosmic_sfr}
\end{wrapfigure}

This debate can be addressed with spatially resolved measurements from 3D spectral analysis, allowing to probe various physical quantities, e.g. star formation, gas kinematics, chemical abundances. The MASSIV sample investigates a redshift range unexplored by Integral Field Spectroscopy (IFS), considering that four distinct 3D surveys are preceding it.  IMAGES \citep{2008A&amp;A...484..173P} sample, with 0.4 < z < 0.75, showed that regular rotating disks are quite similar to the local ones and that mergers are playing an important role in galaxy mass assembly. SINS \citep{2009ApJ...706.1364F} sample, \cite{2009ApJ...699..421W} and  \cite{2009ApJ...697.2057L} samples, and LSD/AMAZE \citep{2008A&amp;A...488..463M} sample, with z>1.5 (mostly with z>2), showed that a lot of young galaxies are experiencing a high gaseous turbulence.

The MASSIV sample was built to probe representative galaxies in the redshift range 0.8 < z < 1.9 (Fig. \ref{cosmic_sfr}), where the cosmic star formation rate history is expected to peak \citep{2012A&amp;A...539A..31C}, and where we look forward to a transition between small and disturbed galaxies towards the  Hubble sequence. This redshift range is also an opportunity to infer the establishment of the stark dichotomy within the galaxy population, already in place at z=1.

\begin{wrapfigure}{r}{0.4\textwidth}
	\begin{center}
		\includegraphics[trim=2cm 6cm 2cm 5cm,width=0.40\textwidth]{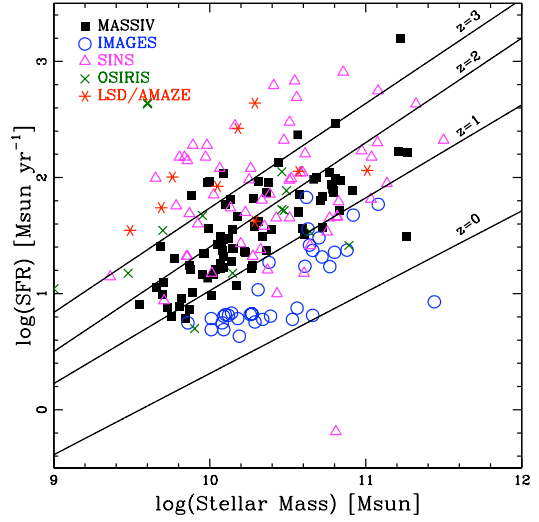}
	\end{center}
	\caption{SED-derived star formation rate as a function of stellar mass. The lines represent the empirical relations between SFR and stellar mass for different redshifts between z = 0 and z = 3 following the analytical expression given in \cite{2010ApJ...718.1001B}. All the illustrated IFU samples are rescaled to match the MASSIV sample redshift range. Adapted from \cite{2012A&amp;A...539A..91C}.}
	\label{sample_comp}
\end{wrapfigure}

MASSIV is an ESO large program (200 hours) grouping 84 galaxies observed with the NIR-IFU SINFONI on the VLT from 2008 until completion in 2011 \citep{2012A&amp;A...539A..91C}. 
The J- or H-bands have been used to target the redshifted H$\alpha$ emission line with a high spatial resolution (<0.8"), and a total integration time varying between 80 and 120 min. 
Among these 84 galaxies, 11 were observed with the adaptive optics system (AO), reaching a spatial resolutions close to 0.20". As of today, it is the largest sample of high-z galaxies observed with IFS.

The strength of this survey lies in its well defined parent sample, that is the VVDS \citep{2005A&amp;A...439..845L} a redshift survey selected in magnitude ($I_{AB}\leq 24$) including 35000 spectra in the visible, avoiding any biases linked to a priori color selection techniques.  A high completeness of the parent sample is mandatory if one wants to probe normal and representative galaxies. Galaxies were selected on [OII]3727 equivalent width strength or rest-frame UV intensity from SED fitting, both being a proxy for star formation. This star formation criterion ensures that the brightest rest-frame optical emission line H$\alpha$  ([OIII]5007 for a few galaxies) is available to probe resolved kinematics and chemical abundances down to galaxies with a SFR close to 1 $M_{\odot}yr^{-1}$.
Galaxies were selected to have a sufficiently close bright star valuable for AO/LGS. The continuum I-band magnitude of each galaxy is estimated using the best-seeing CFHT-LS images with a resolution better than 0.65".

Fig. \ref{sample_comp}  compares the relation between stellar mass and star formation rate for the major IFU samples, over-plotting empirical relations for different redshifts. On the one hand, we see that for a given stellar mass SINS, LSD/AMAZE  are globally probing galaxies with higher star formation rate than MASSIV, while on the other hand, for a given stellar mass, the IMAGES sample is globally probing galaxies with a lower star formation rate than what is expected for a representative sample.

\section{Kinematical analysis of MASSIV galaxies}

\begin{wrapfigure}{r}{0.6\textwidth}
	\centering
	\subfloat[]{\label{fig:high_shear}\includegraphics[trim=0cm 0cm 0cm 1.15cm,width=0.3\textwidth]{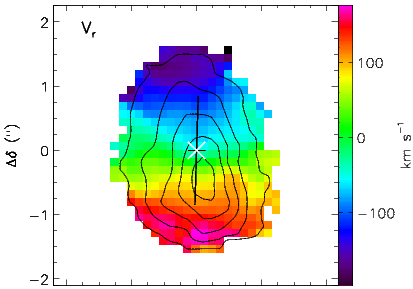}}
	\subfloat[]{\label{fig:low_shear}\includegraphics[trim=0cm 0cm 0cm 1.15cm,width=0.3\textwidth]{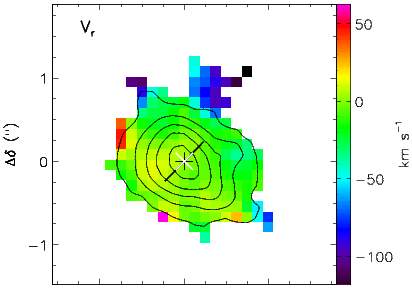}}
	\caption{(a) High-shear velocity field (VVDS220584167). (b) Low-shear velocity field (VVDS020386743). Contours are drawing the H$\alpha$ flux distribution. Adapted from \cite{2012A&amp;A...539A..92E}.}
	\label{low_high_shear}
\end{wrapfigure}

The full MASSIV sample allows resolved velocity measurement for $\sim$90\% of the galaxies. Each velocity field was fitted using a PSF-convolved 'flat' model rotation curve \citep{2010MNRAS.401.2113E}. A kinematical classification of the first epoch sample has been performed in \cite{2012A&amp;A...539A..92E}, using three distinct estimators. A first parameter used to distinguish fast rotators from slow rotators is the total velocity shear $V_{shear}$ measured on the velocity field, without any inclination correction (see Fig. \ref{low_high_shear}). With this simple parameter applied on the full sample,  33 (42\%) galaxies exhibit high velocity shear ($V_{shear}$>100 $km.s^{-1}$), and 45 (58\%) galaxies have a low velocity shear ($V_{shear}$<100 $km.s^{-1}$). 

We also compared a parameter measuring the discrepancy between the position angle of the major axis (PA) of the stellar component and the kinematical PA of the gaseous component with the mean weighted velocity field residuals normalized by the velocity shear. Similary it gives 33 (42\%) galaxies classified as "rotating", and 45 (58\%) galaxies classified as "non-rotating" systems.

In addition, the immediate environment of each galaxy is probed to determine whether the galaxy lies isolated or not. We search for companions in both the H$\alpha$ and I-band image, with a relative velocity lower than 1000 $km.s^{-1}$ and a projected distance lower than 50 $h^{-1}.kpc$. Applying this method, 59 (76\%) galaxies are found to be isolated, and 19 (24\%) are not. 

This work on close environment has been pushed further away by \cite{2012arXiv1208.5020L}. Defining a close pair as a couple of galaxies with a projected radial separation lower than 20 $h^{-1}.kpc$, and a radial velocity difference lower than 500 $km.s^{-1}$,  we are able to recover a major merger rate $R_{MM} \propto (1+z)^{3.95}$ in the MASSIV redshift range. 

Although the origin of galaxies with high $V_{shear}$ is clear (disks with ordered rotation), galaxies with low $V_{shear}$ are more difficult to interpret. Face-on disks, unrelaxed merger remnants, or star-forming spheroids could be an explanation to such a behavior. Nevertheless, the orientation of disks with spin vectors randomly distributed could only account for 14\% of this low-shear population considering disks rotating at $V_{rot}$=200 $km.s^{-1}$. The large fraction of interacting, and non-rotating galaxies seems to suggest that an important cosmological mass assembly mechanism is at work between redshift 1 and 2.

\section{Fundamental relations with MASSIV}

In \cite{2012A&amp;A...546A.118V} we investigate the fundamental relations
using MASSIV data. With dynamical arguments, we derive a gas mass that
is on average a fraction of $\sim 45\%$ of the dynamical mass, assuming no
central contribution of dark matter, as in \cite{2011A&amp;A...528A..88G}, and consistent with lower concentration parameter of haloes at high-z \citep{2001MNRAS.321..559B}. This
gas content is consistent with the fraction derived using the
Kennicutt-Schmidt formulation.

\begin{figure}[h]
	\centering
	\includegraphics[width=0.45\textwidth]{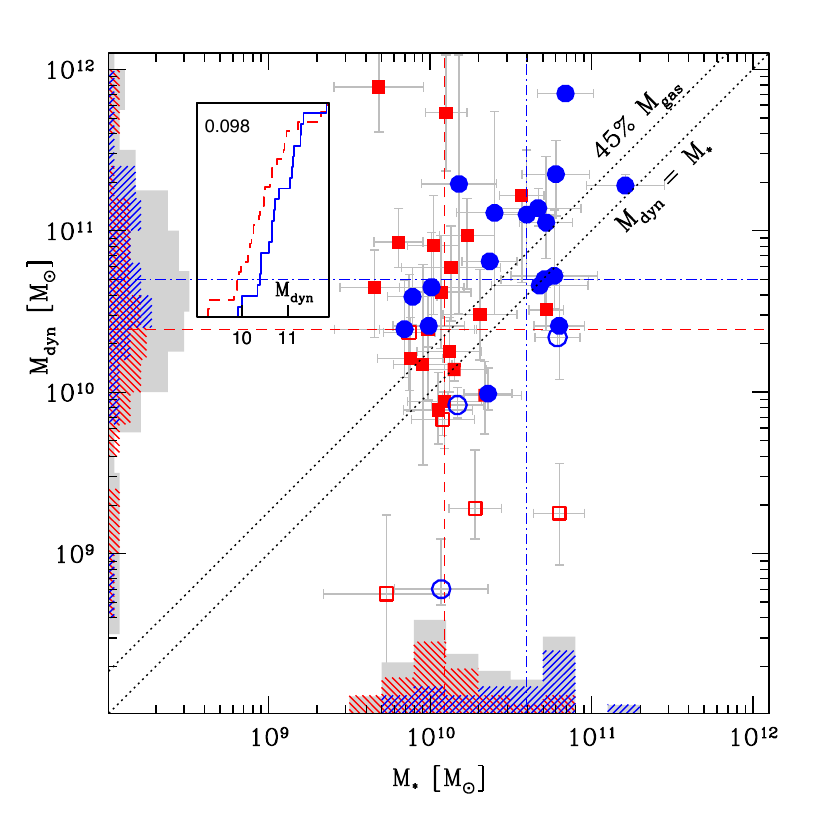}
	\includegraphics[trim=0cm 0cm 0cm 0cm,width=0.45\textwidth]{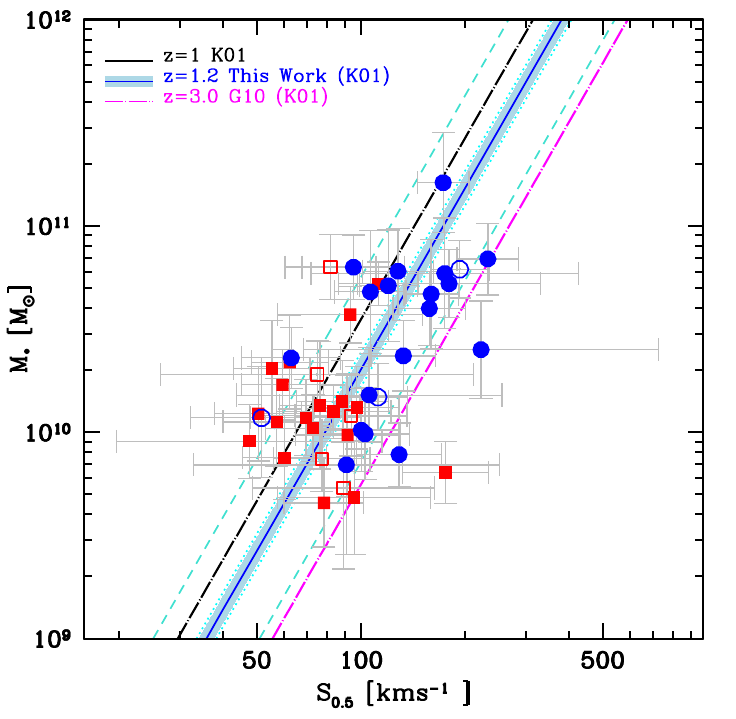}
	
	\caption{\textbf{Left}: Stellar mass content of the MASSIV galaxies compared to the dynamical modeling $M_{dyn}$. The blue/red points corresponds to rotating/non-rotating galaxies according to \cite{2012A&amp;A...539A..92E} criterion. The empty, dashed symbols show galaxies detected with a low signal-to-noise ratio (3 $\leq$ SNR $\leq$4.5). The horizontal and vertical lines are the median values of rotators (point-dashed blue line) and non-rotators (dashed red line). The relation $M_{dyn} = M_{\star} + M_{gas}$ is over-plotted with the mean value of the gas fraction (45\%) for both rotators and non-rotators. \textbf{Right}: Stellar mass TFR at z$\sim$1.2 (blue line) based on rotating MASSIV galaxies (blue dots), taking into account both the ordered and chaotic motions (Kassin et al. 2001, K01). We over-plot the $z\sim1$ slope by \cite{2007ApJ...660L..35K} (solid black line), and the $z\sim3$ slope by \cite{2011A&amp;A...528A..88G} (long-dashed magenta line). The cyan shaded area shows the 1$\sigma$ error on the zero-point parameter, the cyan dotted line and cyan dashed line are the intrinsic and total scatter, respectively. Adapted from \cite{2012A&amp;A...546A.118V}.}
	\label{stellar_dynamical_mass}
\end{figure}

The evolution of the Tully-Fischer Relation (TFR) is expected to be
related both to the conversion of gas into stars and to the inside-out
growth of dark matter halo by accretion. In \cite{2012A&amp;A...546A.118V}
the stellar mass TFR shows a negligible, net evolution in the past 8~Gyrs with a
large scatter that is reduced, but still remarkable, using the
$S_{05}$ index ($S_{05} =\sqrt{0.5 \times v^2_{rot}+\sigma^2_0}$,
Fig. \ref{stellar_dynamical_mass}). We interpret this behavior as an evidence of complex
physical mechanism(s) at work in our stellar mass/luminosity regime
and redshift range. We also conclude a marginal evolution in the size -
stellar mass and size - velocity relations in which disks become
evenly smaller with cosmic time at fixed stellar mass or velocity, and
are less massive at a given velocity than in the local Universe in
agreement with cosmological hydrodynamical simulations, e.g. \cite{2007MNRAS.375..913P}.

\section{Abundance gradients with MASSIV}
 
 The first epoch sample (50 galaxies) enabled an abundances analysis in \cite{2012A&amp;A...539A..93Q}. The metallicity of galaxies was derived in taking the [NII]/H$\alpha$ ratio, with the \cite{2009MNRAS.398..949P} calibration. Among these 50 galaxies, 26 metallicity gradients were measured inside spatially distinct annular regions defined by the H$\alpha$ contour map. While half of the sample (14 galaxies) is compatible with a zero metallicity gradient, a quarter (7 galaxies) of the restricted sample exhibits positive metallicity gradients. 
Among these positive gradients, 4 are classified as interacting systems, one might be a chain galaxy, and 2 are flagged as isolated. Such features have been observed in local interacting galaxies \citep{2010ApJ...715..656W}; it would be an effect of a fresh metal-poor gas infall onto the core of the merger remnant.

The isolated galaxies detected with positive gradients  are more puzzling objects. Such systems have been observed in \cite{2009ApJ...697..115C}, suggesting that cold gas accretion onto the center of the central regions would be a plausible scenario. The radial abundances of the second epoch sample will be published soon, and they will extent the statistics of this first analysis.

\section{Simulations}
High redshift disks have been performed in numerical simulations by \cite{2011ApJ...730....4B} who have shown that the input of large gas fraction (40-60\%) in the initial conditions led to chaotic and turbulent gaseous disks.  Among different question to tackle in the near future, the understanding of the underlying processes responsible for the non-rotating galaxies in the redshift range 0.8 < z < 1.9 led us to simulate high redshift disks from idealized mergers. We are sampling different orbital parameters and mass ratio through a set of 20 simulations.  These simulations will be projected on the sky plane at different evolutionary time using numerous projection angles to mimic observational configurations.  Using the RAMSES code \citep{2002A&amp;A...385..337T}, we are producing simulations of mergers with a physical resolution reaching 6 pc, coupled with a new implementation handling the feedback of OB-type stars emitting energetic ultraviolet photons. With this data we aim to disentangle the different scenarios leading to the formation of a population of galaxies with a low $V_{shear}$, as well as the conditions required for the disk survival.

\vspace{1cm}
\bibliography{buzios_proceeding}

\end{document}